\documentclass[fleqn,twoside]{article}
\usepackage{espcrc2,graphicx}
\usepackage{epsf}

\usepackage{amsmath, amssymb,latexsym,float,algorithm,algorithmic}
\usepackage{enumerate}

\newtheorem{thm}{Theorem}[section]

\newtheorem{prop}[thm]{Proposition}

\numberwithin{equation}{section}

\newcommand{\Real}{\mathbb R}
\newcommand{\Complex}{\mathbb C}

\newcommand{\R}{\text{\fontshape{n}\selectfont I\kern-.42exR}}

\newcommand{\1}{\text{\fontshape{n}\selectfont 1\kern-.56exl}}

\title{The two-grid algorithm confronts a shifted unitary
       orthogonal method}

\author{A. Bori\c{c}i\address[MCSD]{Department of Physics, University of Tirana \\
        Bulevardi Zog I, Tirana-Albania \\
        {\it borici@fshn.edu.al}}%
        }
       
\begin{document}

\begin{abstract}
In this paper I describe a new optimal Krylov subspace solver for shifted unitary
matrices called the Shifted Unitary Orthogonal Method (SUOM).
This algorithm is used as a benchmark against any improvement
like the two-grid algorithm. I use the latter to show that the overlap
operator can be inverted by
successive inversions of the truncated overlap operator.
This strategy results in large gains compared to SUOM.
\vspace{1pc}
\end{abstract}

\maketitle

It is well-known that overlap fermions \cite{Ne98}
lead to much more expensive computations than standard fermions,
i.e. Wilson or Kogut-Sussking fermions. This is obvious since for every
application of the overlap operator an extra linear system solving is needed. 
For the time being, it seems that to get chiral symmetry at finite lattice
spacing one should wait for a Petaflops computer being built.

However, algorithmic research is far from exhausted. In this paper I give an
example that this is the case if one uses the two-grid algorithm \cite{Borici_MG}.
Before I do this, I introduce briefly an optimal Krylov subspace solver for
shifted unitary matrices.

\section{SUOM: A NEW OPTIMAL KRYLOV SOLVER}

Consider the task of solving the linear system:
\begin{equation}\label{lin_sys}
Dx = b, ~~~D = c_1 \1 + c_2 V
\end{equation}
where $V = \gamma_5$sign$(H_W)$ is a unitary matrix, \1 the identity matrix,
$H_W$ the Hermitian Wilson operator, $c_1 = (1 + m_q)/2, c_2 = (1 - m_q)/2$
and $m_q$ the bare fermion mass.
The overlap operator $D$ is non-Hermitian. For such operators 
GMRES (Generalised Minimal Residual) and 
FOM (Full Orthogonalisation Method) are known to be the fastest.
It is shown that when the norm-minimising process of GMRES is converging rapidly,
the residual norms in the corresponding Galerkin process 
of FOM exhibit similar behaviour \cite{CullumGreenbaum96}.
But they are based on long recurrences and thus require to store
a large number of vectors of the size of matrix columns. However,
exploiting the fact that the overlap operator is a shifted unitary matrix one
can construct a GMRES type algorithm with short recurrences
\cite{JagelsReichel94}.

Similarly, a short recurrences algorithm can be obtained from FOM.
The method is based on an observation of Rutishauser \cite{Rutishauser66}
that for upper Hessenberg unitary matrices one can write $H = L U^{-1}$,
where $L$ and $U$ are lower and upper bidiagonal matrices.
Applying this decomposition for the Arnoldi iteration:
\begin{equation}
V Q_k= Q_k H_k + h_{k+1,k}  q_{k+1} e_k^T
\end{equation}
one obtains an algorithm which constructs Arnoldi
vectors $Q_k$ by short recurrences \cite{Borici_SUOM}:
\begin{equation}
V Q_k U_k = Q_k L_k + l_{k+1,k}  q_{k+1} e_k^T.
\end{equation}
Projecting the linear system (\ref{lin_sys}) onto the Krylov subspace one
gets:
\begin{equation}
(c_1 \1_k + c_2 L_k U_k^{-1}) y_k = e_1
\end{equation}
which can be equivalently written as:
\begin{equation}
(c_1 U_k + c_2 L_k) z_k = e_1, ~~~y_k = U_k z_k.
\end{equation}
Note that the matrix on the left hand side is tridiagonal. It can
be shown that one can solve this system and therefore the original system
using short recurrences \cite{Borici_SUOM}. The resulting algorithm
is called the Shifted Unitary Orthogonal Method (SUOM) and is given below:
\vspace{-0.8cm}
\begin{algorithm}[htp]
\caption{\hspace{.2cm} SUOM algorithm}
\label{suom_algor}
\begin{algorithmic}
\STATE $\rho = ||b||_2; ~q_0 = b/\rho; ~w_0 = q_0$
\STATE $l_{00} = q_0^H V q_0$
\STATE $\tilde{q} = Vq_0 - l_{00} q_0$
\STATE $l_{10} = ||\tilde{q}||_2; ~q_1 = \tilde{q} / l_{10}$
\STATE $\tilde{l}_{00} = c_1 + c_2 l_{00}$
\STATE $\alpha_0 = \rho/\tilde{l}_{00}; ~x_0 = \alpha_0 w_0;
~r_0 = b - \alpha_0 D w_0$
\FOR{$~k = 1,2, \ldots$}
    \STATE $u_{k-1k} = - q_{k-1}^H V q_k / q_{k-1}^H V q_{k-1}$
    \STATE $l_{kk} = q_k^H V q_k + u_{k-1k} q_k^H V q_{k-1}$
    \STATE $\tilde{q} = (V - l_{kk}) q_k + u_{k-1k} V q_{k-1}$
    \STATE $l_{k+1k} = ||\tilde{q}||_2$
    \STATE $q_{k+1} = \tilde{q} / l_{k+1k}$
    \STATE $\tilde{l}_{kk} = c_1 + c_2 l_{kk}
- c_1 c_2 l_{kk-1} u_{k-1k} / \tilde{l}_{k-1k-1}$
    \STATE $\alpha_k = - c_2 l_{kk-1} / \tilde{l}_{kk} \alpha_{k-1}$
    \STATE $w_k = q_k + u_{k-1k} q_{k-1}
 - c_1 u_{k-1k} / \tilde{l}_{k-1k-1} w_{k-1}$
    \STATE $x_k = x_{k-1} + \alpha_k w_k$
    \STATE $r_k = r_{k-1} - \alpha_k D w_k$
    \STATE Stop if $||r_k||_2 <$ tol $\rho$
\ENDFOR
\end{algorithmic}
\end{algorithm}
\vspace{-0.9cm}

Note that in an actual implementation one can store $Vq_k$ and $Dw_k$
as separate vectors, which can be used in the subsequent iteration to
compute $Dw_{k+1}$.
Therefore only one multiplication by $V$ is needed at each step.

\section{THE TWO-GRID ALGORITHM}

A straightforward application of multigrid algorithms is hopeless
in the presence of non-smooth gauge fields. However, the situation
is different for the 5-dimensional formulation of chiral fermions
where there are no gauge connections along the fifth dimension.
Here, I will limit my discussion in the easiest case which consists
of two grids: the ``fine'' grid, which is the continuum along the fifth
coordinate and a coarse grid, which is the lattice discretisation of
the ``fine'' grid.

I define chiral fermions on the coarse grid using truncated overlap fermions
\cite{Borici_TOV}. The corresponding 5-dimensional matrix ${\mathcal M}$ in
blocked form is given by:
\begin{equation*}
\hspace{-0.2cm}
\begin{small}
\begin{pmatrix}
D_W-\1          & (D_W+\1)P_+ & -m_q(D_W+\1)P_- \\
(D_W+\1)P_-     & \ddots      &                 \\
                & \ddots      & \ddots          \\
-m_q(D_W+\1)P_+ &             & D_W-\1          \\
\end{pmatrix}
\end{small}
\end{equation*}
where $P_{\pm} = (\1_4 + \gamma_5)/2$. Let ${\mathcal M}_1$ be the above matrix
but with bare quark mass $m_q = 1$ and $P$ the permutation matrix:
\begin{equation*}
\begin{small}
\hspace{1cm}
\begin{pmatrix} P_+ & P_- &        &     \\
                    & P_+ & \ddots &     \\
                    &     & \ddots & P_- \\
                P_- &     &        & P_+ \\
\end{pmatrix}
\end{small}
\end{equation*}
It can be shown that the following result hold \cite{Borici_MG,Borici_link}:
\begin{prop}\label{proposition}
Let $P^T {\mathcal M}_1^{-1} {\mathcal M} P \chi = \eta$
be the linear system defined on the 5-dimensional lattice
with $\chi = (y,\chi^{(2)},\ldots,\chi^{(N_5)})^T$ and $\eta = (r,o,\ldots,o)^T$.
Then $y$ is the solution of the linear system $D^{(N_5)} y = r$,
where $D^{(N_5)} \rightarrow D$ as $N_5 \rightarrow \infty$.
\end{prop}
This result lends itself to a special two-grid algorithm
\cite{Borici_MG,Borici_link}. Indeed, $x_5 = a_5$ is the (fifth Euclidean)
coordinate of interest since it contains the information about the
4-dimensional physics.
\vspace{-0.7cm}
\begin{algorithm}[htp]
\caption{\hspace{.2cm} The two-grid algorithm}
\label{two_grid_algor}
\begin{algorithmic}
\STATE $x_1 \in \Complex^N; ~r_1 = b - D x_1;
~\text{tol}, \text{tol}_0 \in \Real_+$
\FOR{$~i = 1,2, \ldots$}
    \STATE Let $\eta_i = (r_i,o,\ldots,o)^T \in \Complex^{NN_5}$
    \STATE Let $\chi_{i+1} =
(y_{i+1},\chi_{i+1}^{(2)},\ldots,\chi_{i+1}^{(N_5)})^T \in \Complex^{NN_5}$
    \STATE Solve ~${\mathcal M} P\chi_{i+1} = {\mathcal M}_1 P\eta_i$ until
    \STATE $||{\mathcal M}_1 P\eta_i - {\mathcal M} P\chi_{i+1}||_2
< \text{tol}_0 ~||{\mathcal M}_1 P\eta_i||_2$
    \STATE $x_{i+1} = x_i + y_{i+1}$
    \STATE $r_{i+1} = b - D x_{i+1}$
    \STATE Stop if $||r_{i+1}||_2 <$ tol $||b||_2$
\ENDFOR
\end{algorithmic}
\end{algorithm}
\vspace{-0.8cm}
One way of exploiting this is to use ``decimation''
over the fifth coordinate in order to get the 5d-vector $\eta$.
Using proposition \ref{proposition} one can evaluate directly
the first 4d-component of $\eta$ by $r = b - D x$,
$x$ being an approximate solution. The rest can be padded with zero 4d-vectors.
The second step is to solve the problem on the coarse grid.
Finally, one can extract the 4d-solution $y$ on this grid and correct the
``fine'' grid solution by $x \leftarrow x + y$. In the second cycle one has
to repeat the same decimation method, since the ``fine'' 5d-operator is
not available. Hence, the whole scheme is a restarted two-grid algorithm,
which is given here as Algorithm \ref{two_grid_algor}.

\section{COMPARISON OF METHODS}

In Fig. 1 is shown the convergence of variuos algorithms as a function of
Wilson matrix-vector multiplication number on a fixed gauge background
on a $8^316$ lattice at $\beta = 5.7$. The convergence is measured using
the norm of the residual error. For the overlap matrix-vector multiplication
is used the double pass Lanczos algorithm (without small eigenspace
projection of $H_W$) as described in \cite{Borici_over}. Together with the
algorithms described in the previous sections Fig. 1 shows the performance
of Conjugate Residuals (CR), Conjugate Gradients on Normal Equations (CGNE)
and CG-CHI. The latter is the CGNE which solves simultaneously the decoupled
chiral systems appearing in the matrix $D^HD$. One can observe a gain over
CGNE which may be explained due to the reduced number of eigenvalues at each
chiral sector. However, this gain is no more than $10\%$. On the other hand
SUOM and CR preform rather similar with SUOM being slightly faster in this scale.
The gain over CGNE is about a factor two. The Two-Grid algorithm performs the
best with a gain of at least a factor 6 over SUOM and more than an order of
magnitude over CGNE. This situation repeats itself for a different gauge
configuration which is not shown here for the lack of space. However, if the
projection of small eigenvalues is used the gain over SUOM/CR should be smaller
since the Two-Grid algorithm is much less intensive in the application of
the overlap operator. It is exactly the purpose of this comparison to make clear
this feature of the Two-Grid algorithm. Finally, it is (not) surprising
that SUOM and CR perform similarly: CR can be shown to be an efficient method
for normal matrices. Since it is easier to implement
CR is more appealing than other Krylov solvers.

The results of this work suggest that the full multigrid algorithm
along the fifth dimension should be the next method to be explored.

{\bf Aknowledgements.} The author would like to thank Andreas Kronfeld and
LOC of Lattice 2004 for the kind support and Soros Foundation Albania for granting
the travel to Fermilab.

\begin{figure}
\epsfxsize=4cm
\vspace{1.5cm}
\hspace{1.8cm} \epsffile[200 400 480 450]{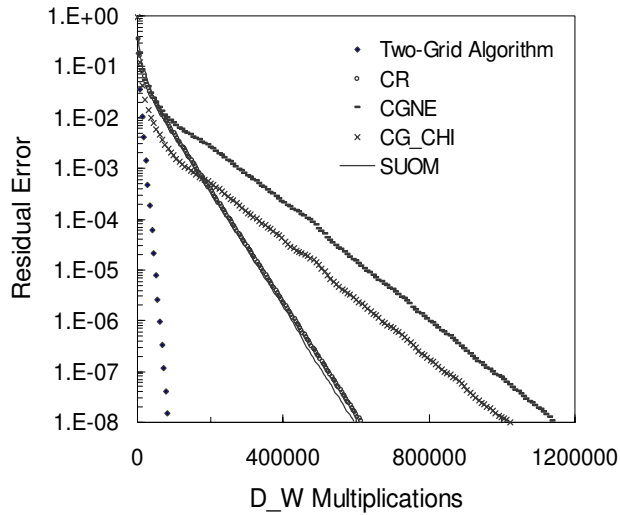}
\vspace{3.6cm}
\caption{Norm of the residual as a function of the number of $D_W(M)$
multiplications for different algorithms for $M = -1.8$}
\vspace{-0.6cm}
\end{figure}


\begin{thebibliography}{9}

\bibitem{Ne98}
        H. Neuberger,
        Phys. Lett. B 417 (1998) 141

\bibitem{Borici_MG}
         A. Bori\c{c}i,
         Phys. Rev. D62 (2000) 017505

\bibitem{CullumGreenbaum96}
         J. Cullum and A. Greenbaum, 
         SIAM J. Matrix Anal. Appl., V17, 2, pp. 223-247 (1996)

\bibitem{JagelsReichel94}
         C. F. Jagels und L. Reichel,
         Num. Lin. Algeb. Appl., Vol. 1(6), 555-570 (1994)
         and
         G. Arnold {\it et al},
         {\tt hep-lat/0311025}

\bibitem{Rutishauser66}
         H. Rutishauser,
         Numer. Math. 9 (1966) 104

\bibitem{Borici_SUOM}
         A. Bori\c{c}i,
         in preparation.

\bibitem{Borici_TOV}
         A. Bori\c{c}i,
         Nucl. Phys. Proc. Suppl. 83 (2000) 771-773

\bibitem{Borici_link}
        A. Bori\c{c}i,
        in V. Mitrjushkin and G. Schierholz (edts.),
        Lattice Fermions and Structure of the Vacuum,
        Kluwer Academic Publishers, 2000.
        See also
        A. Bori\c{c}i,
        {\tt hep-lat/0402035}.

\bibitem{Borici_over}
        A. Bori\c{c}i,
        Phys. Lett. B453 (1999) 46-53

\end{thebibliography}
\end{document}